\documentclass[10pt,twocolumn,prl,nofootinbib,preprintnumbers,amssymb,amsfonts,amsmath,superscriptaddress,showpacs]{revtex4-1}

\usepackage{graphicx}
\usepackage{bm}% bold math
\usepackage{hyperref}
\usepackage{color}
\newcommand{\eV}{{\rm eV}}

\newcommand{\MeV}{{\rm MeV}}
\newcommand{\GeV}{{\rm GeV}}

\newcommand{\Mpl}{M_{\rm Pl}}

\begin{document}

\title{Ultra high energy photons as probes of \\ Lorentz symmetry violations in stringy space-time foam models}

 \author{Luca Maccione}
 \email{luca.maccione@desy.de}
 \affiliation{DESY, Theory Group, Notkestra{\ss}e 85, D-22607 Hamburg, Germany}
 
 \author{Stefano Liberati}
 \email{liberati@sissa.it}
 \affiliation{SISSA, Via Bonomea 265, 34136 Trieste, Italy}
 \affiliation{INFN, Sezione di Trieste, Via Valerio, 2, I-34127, Trieste, Italy}
 
 \author{G\"unter Sigl}
 \email{guenter.sigl@desy.de}
 \affiliation{II. Institut f\"ur Theoretische Physik, Universit{\"a}t Hamburg, Luruper Chaussee 149, D-22761 Hamburg, Germany}
 
 \preprint{DESY 10-039}

\begin{abstract}
The time delays between gamma-rays of different energies from extragalactic sources have often been used to probe quantum gravity models in which Lorentz symmetry is violated. It has been claimed that these time delays can be explained by or at least put the strongest available constraints on quantum gravity scenarios that cannot be cast within an effective field theory framework, such as the space-time foam, D-brane model. Here we show that  this model would predict too many photons in the ultra-high energy cosmic ray flux to be consistent with observations. The resulting constraints on the space-time foam model are much stronger than limits from time delays  and allow for Lorentz violation effects way too small for explaining the observed time delays.
\end{abstract}
\pacs{98.70.Sa, 04.60.-m, 96.50.sb, 11.30.Cp}

\maketitle

%\section{Introduction}

\noindent{\underline{\em Introduction}:} Recent years have witnessed a growing interest in possible small deviations from the exact local Lorentz Invariance (LI) of general relativity. On the theoretical side, ideas stemming from the Quantum Gravity (QG) community led to conjecture that LI may not be an exact local symmetry of the vacuum. On the observational side, high energy astrophysics observations played a leading role in constraining such models and in particular  the recent detection of time delays on arrival of high energy $\gamma$-rays \cite{Abdo:2009zz,Abdo:2009na} led to renewed interest of the astrophysics community in QG induced Lorentz violation (LV) effects. For a comprehensive review see e.g.~\cite{Mattingly:2005re,Jacobson:2005bg, AmelinoCamelia:2008qg,Liberati:2009pf}. 

The observed time delays can be explained, and are actually expected, in standard astrophysical scenarios hence they can be readily used to place constraints on LV models. However, time delays are naturally predicted also in generic LV QG models. %Hence a vibrant interest 
%(even beyond the scientific community) 
%has been raised by the possibility that some of these models could account for all or part of the observed phenomenology. 
Hence the possibility that QG effects could account for all or part of the observed phenomenology raised a vibrant interest.

It is now established that any LV model able to reproduce the observed delays and admitting an Effective Field Theory (EFT) formulation is in tension with other astrophysical observations (see e.g.~\cite{Liberati:2009pf}). Up to now, the only fully developed LV model able to explain the observed time delays has a string theory origin and does not admit an EFT formulation \cite{Ellis:1992eh,Ellis:2000sf,Ellis:2003if,Ellis:2003sd,Ellis:2008gg, Ellis:2009yx, Li:2009tt, Ellis:2009vq}.\footnote{We shall not consider here the possible alternative of deformed special relativity models  \cite{AmelinoCamelia:2000mn,AmelinoCamelia:2008qg} as their dynamics is not fully developed to date.}
In particular, in \cite{Ellis:2009vq} it is not only suggested that the considered QG model could possibly account for all the observed time delays, but it is shown that it might also explain consistently the dark energy content of the Universe.

Therefore, if observed time delays were due to such QG effects, the propagation of GeV photons over cosmological distances could not be described within EFT. Given that EFT is accurately verified with terrestrial accelerators up to $\sim 100~\GeV$, this would be a very striking and revolutionary conclusion that calls for extra investigations of stringy QG models.

In this Letter we show that experimental data on the photon content of Ultra-High-Energy Cosmic Rays (UHECR) lead, for the first time, to strong constraints on this D-brane LV model, making it unsuitable to consistently explain the observed time delays and probably unnatural from a theoretical point of view. As we will argue in the following, due to suppression of UHE photon absorption on intergalactic radiation fields, the fraction of photons in UHECRs predicted within this model would be so large as to violate present experimental limits.

\noindent{\underline{\em Time delays}:}  Effects suppressed by the Planck scale $\Mpl = 1.22\cdot 10^{19}~\GeV$ are in principle hard to detect. Yet in some peculiar situations these tiny effects can be possibly magnified and become sizable. In order to identify these situations, it is required to work in a well defined theoretical framework to describe particle dynamics.

In the model \cite{Ellis:1992eh,Ellis:2000sf,Ellis:2003if,Ellis:2003sd,Ellis:2008gg, Ellis:2009yx, Li:2009tt, Ellis:2009vq} only purely neutral particles, such as photons or Majorana neutrinos, possess LV modified dispersion relations. For photons this has the form 
\begin{equation}
E_{\gamma}^{2} = p^{2} - \xi \frac{p^{\alpha}}{M^{\alpha-2}}\;,
\label{eq:mdr}
\end{equation}
with the free parameter $\xi>0$. Hence only subluminal photons are present in the theory, and photon propagation in vacuum is not birefringent. % (indeed, the theory preserves CPT). %\footnote{In the case of EFT, the dimension 5 operators leading to a modified dispersion relation with $\alpha = 3$ violate CPT, and in particular the two opposite circular photon polarization states have also opposite values of $\xi$ \cite{Jacobson:2005bg}, meaning that their propagation in vacuum is birefringent. The case $\alpha = 4$ in EFT is instead CPT invariant, hence non birefringent.} 
In particular, the model outlined in \cite{Ellis:1992eh, Ellis:2000sf, Ellis:2003if, Ellis:2003sd, Ellis:2008gg, Ellis:2009yx, Li:2009tt, Ellis:2009vq} predicts $\alpha = 3$, hence we will fix $\alpha = 3$ in the following.
Due to stochastic losses in interactions with the D-brane foam, exact energy-momentum conservation during interactions does not hold. 
%This is possible if interactions with the D-brane foam have much shorter timescales than particle interactions. 
This last phenomenon is controlled by the free parameter $\xi_{I}$ \cite{Ellis:2000sf}, in a way which we will clarify below. Because both $\xi$ and $\xi_{I}$ are dimensionless, their natural values are $\mathcal{O}(1)$, and constraints stronger than $\mathcal{O}(1)$ mean that extra suppression of the LV effects has to be invoked.

This model evades most of the present constraints. The electron and birefringence constraints discussed in \cite{Liberati:2009pf} do not apply, because the theory has LV only in the photon (and Majorana neutrino) sector, it is not birefringent, and LV applies only to real (on shell) particles \cite{Li:2009tt}. UHECR constraints \cite{Aloisio:2000cm,Maccione:2009ju,Stecker:2004xm,Scully:2008jp,Stecker:2009hj,GonzalezMestres:2009di} do not apply as well. 

However, according to Eq.~(\ref{eq:mdr}) photons with different energy travel at different speeds. Then, if a source at redshift $\bar{z}$ simultaneously emitted two photons at energy $E_1' \neq E_2'$, their time delay at Earth will be
\begin{equation}
\Delta t \simeq \xi \frac{\Delta E}{M}\frac{1}{H_{0}}\int_{0}^{\bar{z}}dz\frac{1+z}{\sqrt{\Omega_{\Lambda} + (1+z)^{3}\Omega_{\rm M}}}\;,
\label{eq:tof}
\end{equation}
where $\Delta E$ is the observed energy difference and the integral on redshift accounts also for redshift of the energy \cite{Jacob:2008bw,AmelinoCamelia:2009pg,Ellis:1999sd}. Time-of-flight constraints are then viable for this model, even though they lead at most to constraints on $\xi$, because $\xi_{I}$ is not effective in this context. 

Rather intriguingly, the FERMI Collaboration has recently reported the detection of delays on arrival of $\gamma$-ray photons emitted by distant GRBs, in particular GRB 080916C \cite{Abdo:2009zz} and GRB 090510 \cite{Abdo:2009na} (see however \cite{AmelinoCamelia:2009pg} for an updated review). A thorough analysis of these delays in the energy range 35 MeV -- 31 GeV allowed to place for the first time a conservative constraint of order $\xi \lesssim 0.8$ \cite{Abdo:2009na} on LV effects expressed as in Eq.~(\ref{eq:mdr}). 
This is the best constraint so far available on the theory. On the other hand, FERMI results can be interpreted in terms of LV assuming $\xi\simeq 0.4$ and a possible evolution of the D-particle density with redshift~\cite{Ellis:2009vq}.\footnote{Plausible astrophysical explanations of this phenomenon exist. No claim of a discovery of LV can be made on the basis of the data reported in \cite{Abdo:2009zz, Abdo:2009na}, where only LV constraints are discussed.}

\noindent{\underline{\em Photon absorption in D-brane models}:} In order to constrain the D-brane model, we exploit the process of pair production, $\gamma\gamma \rightarrow e^{+}e^{-}$, which is in particular responsible for the absorption of UHE photons produced in GZK interactions \cite{Greisen:1966jv,Zatsepin:1966jv}. Indeed, if GZK energy losses affect the propagation of UHECR protons in the intergalactic medium, then a large amount of UHE photons is generated by the decay of the $\pi^0$'s copiously produced in such interactions. UHE photons are attenuated by pair production onto the CMB and Radio background during their travel to Earth, leading to their fraction in the total UHECR flux being reduced to less than 1\% at $10^{19}\eV$ and less than 10\% at $10^{20}~\eV$ \cite{Sigl:2007ea, Gelmini:2007jy}. It was shown in a framework with modified dispersion relations for both photons and $e^{+}/e^{-}$  and standard energy/momentum conservation, that pair production could be effectively inhibited at high energy, due to the presence of an upper threshold \cite{Galaverni:2007tq},\footnote{An upper threshold is an energy above which it is not possible to simultaneously conserve energy and momentum in an interaction.  If Lorentz symmetry is exact then upper thresholds do not exist, while they might well exist if it is violated \cite{Mattingly:2002ba}.} and therefore the fraction of photons present in UHECRs on Earth would violate the present experimental upper limits. Hence, the {\em non} observation of a large fraction of UHE photons in UHECRs implies the constraint $|\xi| < \mathcal{O}(10^{-14})$ in the EFT framework \cite{Maccione:2008iw,Galaverni:2008yj}. 

We want to address here the problem whether the same argument can be applied in the space-time foam model with energy non conservation.
%%%%%
First, we note that $\gamma$-rays are indeed generated by $\pi^{0}$ decay, i.e.~that $\pi^{0}$ decay is not affected by LV in our working scenario: In D-brane models LV acts as an energy dependent modification of the background space-time metric (the new metric is of Finslerian type \cite{Ellis:1999uh,Li:2009tt}), hence it can act only on real particles. But in order for $\pi^{0}$s to decay, they must excite modes of the electromagnetic vacuum, i.e.~virtual photons, which then are not affected by LV. Hence $\pi^{0}$ decay is not affected by LV. The LV effects described in D-brane models are the result of multiple interactions between photons and D-branes, hence cannot be relevant in the mere process of photon production.\footnote{In fact, this seems to justify a more general theorem, stating that only reactions with photons in the {\em initial} state can be affected by LV in this model. Note that the resulting violation of time reversal invariance
is due to LV in this scenario. We thank N.~Mavromatos for bringing this argument to our attention.} Moreover, $\pi^{0}$ is not a structureless particle, and its constituents are charged, hence they do not interact with D-particles. 

%The dynamics of pion decay might in general be affected by the presence of D-branes, which introduces an upper threshold $p_{\pi}^{\rm up\; th} = \left(m_{\pi}^{2}\Mpl\right/(\xi_{I}-\xi))^{1/3}$. This means that for $\xi_{I} \leq \xi$ the pion decay proceeds almost unaffected (the correction to the photon energy due to LV is negligible, even at $10^{20}~\eV$), while in the opposite case $\xi < \xi_{I}$ there is an upper threshold and UHE pions may not decay. The upper threshold is found at $10^{19}~\eV$ for $\xi_{I}- \xi \simeq 10^{-13}$. 
%%%%%%%%%%
Now we consider the threshold equations \cite{Ellis:2000sf}
\begin{eqnarray}
\nonumber
E_{1}+\omega &=& E_{2}+E_{3} + \delta E_{D}\\
p_{1}-\omega &=& p_{2}+p_{3}\;,
\end{eqnarray}
where $\omega$ is the energy of the low energy background photon ($\omega \simeq 6\times 10^{-4}~\eV$ for a CMB photon), $E_{1} \simeq p_{1} -\xi/M\cdot p_{1}^{2}/2$ is the energy of the high energy photon and $E_{j} \simeq p_{j}+m_{e}^{2}/(2p_{j})$, with $j=2,3$ are the energies of the outgoing electron and positron. The symbol $\delta E_{D}$ represents the energy lost in the stochastic interactions with the D-branes. The above equation is already written in the threshold configuration %(head-on collision and collinear outgoing particle momenta)
 \cite{Mattingly:2002ba}. We exploit momentum conservation and the ultrarelativistic limit, to get %(we use that $p_{1} = \omega + p_{2}+p_{3}$)
\begin{equation}
2\omega - \frac{m_{e}^{2}}{2p_{2}} - \frac{m_{e}^{2}}{2p_{3}} = \delta E_{D}^{(4)} + \frac{\xi}{M}\left(\omega^{2}+\omega(p_{2}+p_{3})+p_{2}p_{3}\right)
\end{equation}
where $\delta E_{D}^{(4)}$ is the amount of energy violation in a four-particle interaction, and corresponds to the sum of the corresponding violations in each of the two three-body interactions described by Eq.~(27) in \cite{Ellis:2000sf} 
\begin{equation}
\delta E_D^{(4)} \equiv \delta E_D + \frac{\xi}{2M}\left(p_2^2 + p_3^2 -\omega^2\right) \simeq \frac{\xi_{I}}{2M}E_{\rm th}^{2}\;,
\label{eq:deltaED4}
\end{equation}
where according to \cite{Ellis:2000sf} we assume the last equality to hold, with $\xi_{I}$ different from $\xi$ in principle. While it is natural that $\delta E_{D}^{(4)}$ depends only on $E_{\rm th}$ and $M$, as they are the only energy scales of the problem, the effect of the quantum fluctuations $\delta E_{D}$ is less clear. We shall assume that this effect only amounts to a redefinition of the unknown parameter $\xi_{I}$, and check that our conclusions remain unchanged if we let $\xi_{I}$ fluctuate in the interaction up to 10 times its central value. 
The threshold equation can then be derived by putting $p_{2} = p_{3} = (p_{1}-\omega)/2$ 
%(the threshold configuration of the $e^+/e^-$ momenta is symmetric in this context)
 and $p_{1} \equiv E_{\rm th}$. By introducing $x \equiv E_{\rm th}/M$ we obtain
\begin{widetext}
\begin{equation}
-\frac{\xi_{I}+\xi/2}{2}x^{3} + \frac{\xi_{I}-\xi/2}{2}\frac{\omega}{M}x^{2} + \left(2+\frac{\xi}{4}\frac{\omega}{M}\right)\frac{\omega}{M}x - 2\frac{\omega^{2}+m_{e}^{2}}{M^{2}}+\frac{\xi}{4}\left(\frac{\omega}{M}\right)^{3} = 0\;.
\end{equation}
\end{widetext}
%
%where $x \equiv E_{\rm th}/M$.

\begin{figure}[tbp]
\centering
\includegraphics[scale = 0.38]{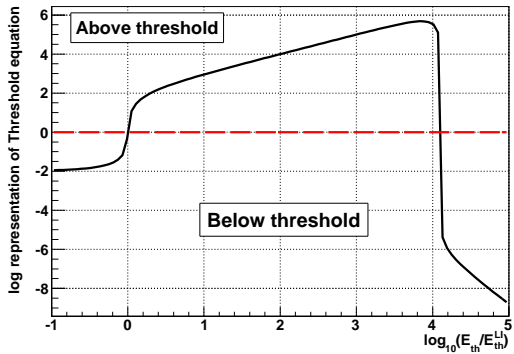}
% threshold_eq.eps: 0x0 pixel, 300dpi, 0.00x0.00 cm, bb= 0 0 567 384
\caption{Equation (\ref{eq:threshold}) is drawn for $\xi/2+\xi_{I} = 10^{-12}$. On the $y$-axis ${\rm sign}({\rm Eq.~}(\ref{eq:threshold}))\cdot \log_{10}\left(|{\rm Eq.~}(\ref{eq:threshold})|/0.01\right)$ is drawn, where 0.01 is just a small value. Positive values on the $y$-axis mean that the reaction is allowed, while negative values mean that it is forbidden. Both lower and upper thresholds are present. %Of course Eq.~(\ref{eq:threshold}) can be solved for any particular value of $\xi/2+\xi_{I}$.
}
\label{fig:threshold}
\end{figure}

\noindent{\underline{\em Constraints}:} If we now neglect all the terms more than linear in either $\xi$ or $\omega/M$, we recover Eq.~(32) in \cite{Ellis:2000sf}
\begin{equation}
-\frac{\xi_{I}+\xi/2}{2}x^{3} + 2\frac{\omega}{M}x -2\frac{m_{e}^{2}}{M^{2}} + \dots = 0\;.
\label{eq:threshold}
\end{equation}
We represent (up to factors) Eq.~(\ref{eq:threshold}) in Fig.~\ref{fig:threshold}, for $\xi/2 + \xi_I = 10^{-12}$, but we solve it for general values of $\xi/2+\xi_{I}$ to establish constraints.
Equation \eqref{eq:threshold} has in general a lower and an upper threshold ($E_{\rm low}$ and $E_{\rm up}$, respectively). From the observational requirement that  $E_{\rm up}> 10^{19}~\eV$, with $\xi_{I}$ and $\xi$ varying independently %\footnote{Notice however that the structure of the threshold equation implies that the quantity $\Xi = \xi_{I}+\xi/2$ is effectively constrained. However, since $\xi_{I}$ and $\xi$ have the same sign, this translates in an effective constraint on both.}
 (and setting $M=\Mpl$, $\omega = 6\times 10^{-4}~\eV$ and $m_e = 0.511~\MeV$),  values of $\xi_{I},\;\xi> 10^{-12}$ are excluded by the {\em non} observation of a significant photon fraction in the UHECR spectrum by the Auger experiment~\cite{Aglietta:2007yx}. %This is in contrast with the values $\xi \sim O(1)$  being explored by time delay analysis.

It was recently proposed in \cite{Ellis:2009vq} that the evolution with redshift of the D-particle/D-void background might allow to understand why significant delays compatible with $\xi \simeq \mathcal{O}(1)$ are present in some GRBs, while in other GRBs the delays are much smaller and imply $\xi < \mathcal{O}(1)$. Since the effect of time delay (as well as the one of energy non-conservation) is expected to be proportional to the density of D-particles, the scenario envisaged by data requires the density of D-particles to be large for redshift $z > 1$, to drop at $z \sim 1$ and to raise again for $z < 1$. This evolution might in principle affect our constraints, which depend on the effectiveness of pair production at least up to $z \sim 3$. To address this issue, we modified the public UHECR propagation code CRPropa \cite{Armengaud:2006fx}. 

\begin{figure}[tbp]
\centering
\includegraphics[scale = 0.38]{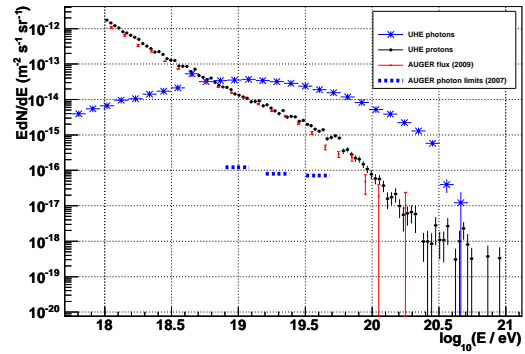}
\caption{UHE proton and photon simulated spectra assuming that pair production is inhibited for $z < 0.2$ and $z > 1$ for a proton injection spectrum $\propto E^{-2.5}$ up to $10^{21}\,$eV and source density redshift evolution as in \cite{Bahcall:2002wi}.
%Proton fluxes are to be read on the left-hand $y$-axis, while photon integrated fluxes $dN/dE(>E_{0})\equiv \int_{E_{0}}^{E_{\rm max}} dE\,dN/dE$ are shown on the right-hand, blue $y$-axis. 
Error bars on the simulated proton flux correspond to the statistical error of the simulation. Measured UHECR flux (in red) is from \cite{Collaboration:2010mj}, while upper limits on the integral photon flux (dashed, in blue) are from \cite{Aglietta:2007yx}. The latter approximates
$E\cdot dN/dE$ within a factor $\sim3$.}
\label{fig:constraints}
\end{figure}

Following \cite{Ellis:2009vq}, we assume that for $0.2 < z < 1$ LV effects are suppressed (then, pair production is allowed as if LI were exact and UHE photons are effectively absorbed), while outside this redshift range we assume that the LV effects are strong and that pair production absorption is inhibited (i.e., we switch it off). Because UHE photons are mainly of local origin, this assumption is conservative: Moving the suppression of LV effects to more distant epochs would indeed increase the photon fraction on Earth. We found that even in this case the photon fraction would violate experimental limits for values of $\xi,\;\xi_I > 10^{-12}$ (see Fig.~\ref{fig:constraints}). %We therefore conclude that the limit is robust against this test.

The interactions between photons and D-particles might be suppressed if the momentum $\Delta p$ transferred to the D-particle is large compared to its mass $M_{D} = M_{s}/g_{s}$, where $M_{s}$ is the string scale and $g_{s}$ is the coupling \cite{Li:2009tt}. In the standard string framework $M_{D}$ is expected to be at least of order $\Mpl$ \cite{Ellis:2009vq}, therefore this would not be an issue for our constraint. However, in some compactification schemes, lower values of $M_{D}$ cannot be excluded \cite{Li:2009tt}. If $\Delta p \gg M_{D}$, $g_{s}$ is replaced by an effective coupling $g_{s}^{\rm eff} = g_{s}/\Gamma$, where $\Gamma \sim \Delta p/M_{D}$, and given that the unknown coefficients $\xi$ and $\xi_{I}$ are proportional to the scattering cross section, which in turn is proportional to $g_{s}^{2}$,  they both receive a natural suppression $1/\Gamma^{2}$. In order to explain the observed time delays in the GeV-TeV energy range within the model \cite{Li:2009tt}, $M_{D}$ has to be substantially larger than the TeV scale. 
%Although $\Delta p$ is not fixed by kinematics, it is possible to estimate its maximum value for the case of two body scattering $\gamma D \rightarrow \gamma D$
However, on the basis of kinematics the maximum suppression factor can be estimated as being $\mathcal{O}(10^{10})$, thereby weakening our constraint to $\xi_{I}, \xi \lesssim 10^{-2}$.
%, finding that, even for $M_{D}\sim 1~\GeV$, $\Delta p \lesssim 10^{-5}E_{{\rm UHE}\gamma}$. Therefore, our constraints are weakened by only a factor $10^{10}$, and we remark that the weakening is less strong with increasing $M_{D}$.

Hence, we conclude that D-particle explanations of GRB time delays, such as in \cite{Ellis:2009vq}, are in conflict with data on the photon fraction in UHECRs.\footnote{After this paper was submitted, and stimulated by its results, some possible implementations of the model \cite{Ellis:2008gg} were recently proposed \cite{Ellis:2010he} which would naturally evade the constraints here presented.}

\begin{acknowledgments}
We thank G.~Amelino-Camelia, D.~Mattingly and especially N.~Mavromatos for useful comments. This work was supported by the Deutsche Forschungsgemeinschaft through the collaborative research centre SFB 676. LM and GS acknowledge support from the State of Hamburg, through the Collaborative Research program ``Connecting Particles with the Cosmos'' within the framework of the LandesExzellenzInitiative (LEXI).
\end{acknowledgments}

%%%%%%%%%%%%%%%%%%%%%%%%%%%%%%%%%%%%%%%%

\end{document}